\documentclass[aps,pra,twocolumn]{revtex4}  
\usepackage{graphicx}
\usepackage{amsmath}    
\usepackage{epstopdf}
\begin{document}

\def\ket#1{|#1\rangle}
\def\bra#1{\langle#1|}
\def\av#1{\langle#1\rangle}
\def\myarrow{\mathop{\longrightarrow}}

\title{Single-photon, cavity-mediated gates: detuning, losses, and non-adiabatic effects}

\author{Julio Gea-Banacloche} 
\affiliation{Department of Physics, University of Arkansas, Fayetteville, AR 72701}
\author{Leno M. Pedrotti} 
\affiliation{Department of Physics, University of Dayton, Dayton, OH 45469}

\date{\today}

\begin{abstract}
We study several extensions of the single-photon, cavity-mediated quantum logical gates recently proposed by Koshino, Ishizaka and Nakamura: to a double-sided cavity configuration, to the case where the two atomic ground states are nondegenerate, and to the non-adiabatic regime.  Our analysis can be used to estimate the effects of various imperfections, and to prepare the way for a proof-of-principle demonstration with present technology.  We are able to present a full solution for the outgoing pulses, in the frequency domain, which is valid for arbitrary incident pulses and cavity parameters.  From this we find, among other results, that the leading correction to the adiabatic approximation can be made to vanish for a suitable choice of detunings, provided the cavity is ``good enough'' (high enough ratio of coupling to loss).  This could significantly relax the need for long single-photon pulses in this scheme. \end{abstract}
\maketitle

\section{Introduction} Photons are widely regarded as natural carriers for quantum information, but their usefulness is limited by the lack of strong photon-photon interactions that could be used to carry out conditional quantum logic on single-photon states.  Although Kerr nonlinearities have often been proposed to address this difficulty, there are good reasons to believe that they cannot really be useful for quantum logic at the single photon level \cite{shapiro,jgbkerr}, at least in an unconstrained geometry \cite{simon}.

Atomic ensembles and optical cavities can be used to constrain the interaction geometry and increase the effective photon-matter coupling, thus providing a possible way to mediate single-photon interactions.  A single-photon gate mediated by a single, three-level atom in a cavity was proposed by Duan and Kimble in 2004 \cite{duan} and generated a great deal of interest, but it has not been experimentally realized yet because of its very challenging nature.  Part of this challenge is the need to address the atom inside the high-finesse optical microcavity with an external laser field (presumably through the sides of the cavity), in order to perform the gate. 

In general, all proposed variations of the Duan-Kimble scheme rely on the different phase shifts acquired by a photon upon reflection from the cavity, depending on whether it is resonant or not with the atomic transition.  Although, as indicated above, a full gate of the Duan-Kimble type has not been demonstrated yet, the required conditional phase shifts have by now been observed in a variety of systems \cite{fushman, Young}, and many potential uses for them in quantum information processing have been suggested (see, for a very small sample, \cite{hu1,an,cheng,hu2}, and references therein).

In contrast to the above proposals, it was shown in 2009 by  Lin et al. \cite{lin} that, if a $\Lambda$-type atom with two \emph{degenerate} transitions (corresponding to different polarizations) in a doubly resonant cavity was used, the incident photon could directly change the atomic state, so that a SWAP gate between the atom and the photon polarization states was possible.  More recently, Koshino, Ishizaka and Nakamura (KIN) showed \cite{kin} that a totally passive, cavity-mediated gate between photons was possible in the same system for an appropriate choice of detunings.  In their scheme, a $\sqrt{\text{SWAP}}$ gate between two single-photon pulses (a two-qubit gate which, together with single qubit gates, suffices to enable universal quantum computation \cite{divincenzo}) could be carried out merely by reflecting the pulses off of the cavity in an appropriate sequence: manipulation of the atom by external fields, whether lasers or static, is not required at all, in principle.  Moreover (as already proposed in \cite{lin}), the KIN system could also be used---again in a totally passive way---as a quantum memory, by swapping the photonic and atomic states.  KIN modules could therefore provide a useful alternative to atomic ensembles, both for storage and manipulation of quantum information.  In a previous paper \cite{prev} we have showed how the KIN scheme (originally proposed for the ``bad'' or ``fast'' cavity case) could be extended to the ``good'' cavity regime.  Other properties of the KIN system, in particular its ability to function as an ideal photon turnstile, have also been recently reported \cite{parkins}.  

In this paper we wish to extend our investigation of the KIN system by relaxing some of the assumptions and restrictions on which previous studies (including our own) have been based, with an aim towards ascertaining the feasibility of a proof-of-principle demonstration with present technology.  The outline of the paper is as follows.

In Section II, for completeness, we present a very brief overview of the original KIN proposal.  In Section III, we consider what happens if one replaces the original single-sided cavity by a double-sided cavity, an important question given that many of the best current optical microcavities are either double-sided or ring cavities (such as microdisks, or microtoroids), and we would like to explore the possibility of using an existing cavity for at least a proof-of-principle experiment (note that some previous proposals, such as \cite{hu2}, naturally work in the double-sided geometry, so it is perhaps not immediately obvious why the KIN scheme would not).  Formally, also, a ``leaky'' second mirror can be used to model the effect of cavity losses other than those arising from the transmission through the first mirror.

In Section IV, we explore the case in which the atomic transitions are not degenerate.  This too is important for several reasons: for example, in some atomic systems the only way to select a closed, lambda-like set of three hyperfine states may be by using external magnetic fields to make all unwanted transitions non-resonant with the frequencies of the photons used.  In such a scheme, the two polarizations making up the photonic qubit would naturally have to have different frequencies as well (as we show in detail below).  Again, alternatively, our results may be taken to show the damaging effect of atomic energy shifts if these are not compensated for.

Lastly, in Section V, we consider the corrections to the adiabatic approximation under which all the KIN results so far have been obtained.  Interestingly, we show that the leading correction to the adiabatic approximation can be made to vanish for a suitable choice of cavity parameters, which could significantly relax the need for long pulses.

\section{The original KIN scheme}

In the original KIN proposal, one considers a single three-level atom, in the $\Lambda$ configuration, inside a single-sided cavity (Fig. 1).  The two ground states are taken to be degenerate, so the two transitions shown correspond to different polarizations.  A single-photon pulse enters the cavity, interacts with the atom, and leaves through the same input mirror.  In the process, it may change the state of the atom or become entangled with it. If the cavity decay rate is $\kappa$, the atomic decay rate is $\gamma$, and the single-photon Rabi frequency (coupling constant to the cavity mode, assumed to be the same for both transitions) is $g$, the ``strong coupling regime'' (assumed throughout) is that $\gamma \ll \kappa, g$.  Within this regime one may still distinguish a ``bad-cavity'' limit, $\kappa \gg g$, and a ``good cavity limit,'' $\kappa \ll g$. In the bad-cavity limit, the cavity parameters enter the equations only through the combination $2g^2/\kappa$, and the strong coupling condition becomes $\gamma \ll 2g^2/\kappa$.

   \begin{figure}
   \includegraphics[width=8cm]{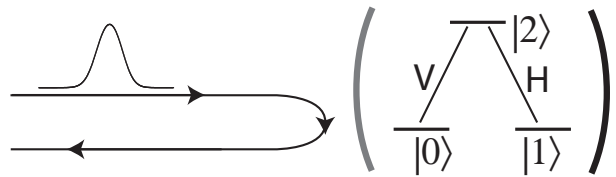}
  \caption[example] 
   { \label{fig:fig1} 
The KIN scheme.  Single-photon pulses are reflected off of a one-sided cavity containing a three-level atom, in the process becoming entangled with the atom.  By swapping the state of a second pulse with that of the atom, an entangling operation between single-photon pulses results.}
   \end{figure} 

In the bad-cavity limit, Koshino, Ishizaka and Nakamura (KIN) showed \cite{kin} that, for a sufficiently long pulse (adiabatic regime), the input and output states of the total atom$+$photon system were related by the transformation
\begin{subequations}
\begin{align}
\ket{H,0} &\to \ket{H,0} \label{n1a}\\
\ket{H,1} &\to -e^{i\phi}\left[i \sin\phi \ket{H,1} + \cos\phi\ket{V,0} \right] \label{n1b}\\
\ket{V,0} &\to -e^{i\phi}\left[\cos\phi \ket{H,1} + i\sin\phi \ket{V,0} \right] \label{n1c}\\
\ket{V,1} &\to \ket{V,1}
\label{n1d}
\end{align}
\label{n1}
\end{subequations}
where $\phi$ is a detuning-dependent parameter.  When $\phi=0$ one has essentially (up to a sign) a SWAP gate between the photon and the atom, whereas when $\phi=\pi/4$ one has a $\sqrt{\text{SWAP}}$ gate \cite{divincenzo}.  To have $\phi = 0$ it is sufficient that all the detunings vanish.  To have $\phi=\pi/4$, Koshino, Ishizaka and Nakamura derived the condition $\delta_a = 2 g^2/\kappa$ for the detuning between the atom and the cavity, assuming implicitly that the cavity and the field were resonant.  For the more general case where the field is detuned from the cavity resonance by an amount $\Delta$, we showed in \cite{prev} that the condition to have $\phi=\pi/4$ is
\begin{equation}
(\Delta+\delta_a)(\kappa^2+\Delta^2)-2 g^2(\Delta+\kappa) = 0
\label{n2}
\end{equation}
This also reduces to the KIN condition in the very bad cavity limit, $\kappa \gg \Delta$.  The above results will be seen to follow as special cases of the more general analysis to be carried out later in this paper.

In their original work, Koshino, Ishizaka and Nakamura studied the effect of some deviations from their assumed ideal setup, such as different coupling constants for the horizontal and vertical polarizations, and finite pulse length.  In \cite{prev}, as stated above, we extended their results to the good cavity limit, considered (in a limited way) the effect of spontaneous emission, and established (numerically) the $1/T^2$ dependence of the gate fidelity on pulse duration $T$.  

In the sections that follow, we shall study, mostly analytically, the effects of other departures from the ideal KIN assumptions: two-sided cavity (or additional cavity losses); nondegenerate ground state; and deviations from the adiabatic evolution (finite-pulse effects).  Our work will be quite self-contained, since we will even present an alternative derivation, and extension, of our ``modes of the universe'' formalism in the following section.

\section{KIN gates with a double-sided cavity}

The ``modes of the universe'' approach was originally introduced by Lang, Scully and Lamb \cite{universe0} as a rigorous way to quantize the quasi-modes of a lossy cavity, such as a laser cavity.  The formal treatment in \cite{universe}, on which our previous paper was based, involves a number of technical subtleties and is not physically very intuitive, so in what follows we present an alternative derivation which is also an extension, to the case of a two-sided cavity with mirrors of (amplitude) transmission coefficients $t_1$ and $t_2$.

The idea is to consider the coupling of an atom inside the cavity to the ``scattering modes'' depicted in Figure 1.  For a given frequency $\Omega_k = ck$ there are two such modes, one incident from the left and one from the right.  For unit incident amplitude, the resulting field amplitudes illustrated in Fig. 2 are, for the mode incident from the left, 
\begin{align}
A &= -\frac{r_1 e^{-i kl} - r_2(r_1^2+t_1^2)e^{ikl}}{r_1 r_2 e^{2ikl}-1}  = -\frac{r_1 e^{-i kl} - r_2 e^{ikl}}{r_1 r_2 e^{2ikl}-1} \cr
B &= - \frac{t_1}{r_1 r_2 e^{2ikl}-1} \cr
C &=   \frac{ t_1 r_2 e^{ikl}}{r_1 r_2 e^{2ikl}-1} \cr
D &=  - \frac{t_1 t_2 }{r_1 r_2 e^{2ikl}-1}
\label{e1}
\end{align}
assuming the cavity has length $l$ and the mirrors have real reflection coefficients, whereas for the mode incident from the right the coefficients $A^\prime$, $B^\prime$, $C^\prime$ and $D^\prime$ are given by identical expressions with the indices $1$ and $2$ switched around.

   \begin{figure}
   \includegraphics[height=4.5cm]{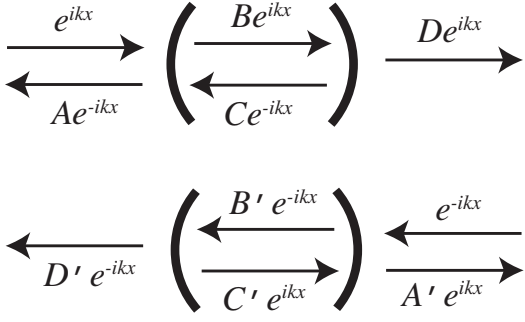}
  \caption[example] 
   { \label{fig:fig2} 
The scattering modes of a double-sided cavity}
   \end{figure} 

Let $a_k$ ($a_{-k}$) be the annihilation operator for the mode incident from the left (right).  At an arbitrary point inside the cavity, an atom would see a field given by
\begin{align}
\sum_k  &\left(\frac{\hbar\Omega_k}{2\epsilon_0 A L}\right)^{1/2} \frac{1}{r_1 r_2 e^{2ikl}-1} \biggl[t_1\left(r_2 e^{ikl}e^{-ikx}-e^{ikx}\right) a_k 
\cr &\qquad+ t_2 \left(r_1 e^{ikl}e^{ikx}-e^{-ikx}\right) a_{-k} \biggr]
\label{e3}
\end{align}
where $AL$ is the quantization volume for the scattering modes.  The resonance condition $kl = (2n + 1) \pi$ minimizes the denominator in (\ref{e3}) while keeping the numerator close to a maximum (for an atom near the center of the cavity, $x\simeq 0$).  We assume then that we can write, for all the relevant modes,
\begin{equation}
e^{2ikl} = e^{2i(\Omega_k-\Omega_c)l/c} \simeq 1+2i(\Omega_k-\Omega_c)l/c
\label{e4}
\end{equation}
where $\Omega_c$ is the cavity's resonant frequency, in the denominator of (\ref{e3}), whereas in the numerator we can simply replace $e^{ikl}$ by $-1$.  The reason for this is that, for a microcavity, $c/l$ is of the order of an optical frequency, and the typical detunings we shall consider will be smaller by many orders of magnitude.  

We can also assume that the reflectivity of the mirrors is very close to 1 and that they are lossless, so $r=\sqrt{1-t^2} \simeq 1 - t^2/2$.  Substituting all this in Eq.(\ref{e3}) we get for the intracavity field
\begin{align}
&\sum_k \left(\frac{\hbar\Omega_k}{2\epsilon_0 A L}\right)^{1/2}  \frac{2 \cos kx}{(t_1^2+t_2^2)/2-2i(\Omega_k-\Omega_c)l/c} \cr
&\qquad \times \left[t_1 a_k + t_2 a_{-k}\right] \cr
&\simeq  \cos k_0 x  \left(\frac{\hbar\Omega_0}{2\epsilon_0 A L}\right)^{1/2}  \sum_k \frac{2\sqrt{\kappa c/l}}{\kappa -i(\Omega_k-\Omega_c)}\left(\tau_1 a_k+\tau_2 a_{-k}\right)\cr
\label{e5}
\end{align}
to lowest order in small quantities.  In the last equality we have replaced $\Omega_k$ by the pulse's central frequency $\Omega_0$ in the ``electric field per photon'' factor.  We have also introduced the total cavity loss rate
\begin{equation}
\kappa = \frac{t_1^2+t_2^2}{4}\,\frac{c}{l}
\label{e6}
\end{equation}
and the parameters 
\begin{equation}
\tau_i = t_i/\sqrt{t_1^2 + t_2^2}\qquad(i=1,2)
\label{e7}
\end{equation}
 satisfying $\tau_1^2 + \tau_2^2 = 1$.    

Suppose one has a two-level atom at $x=0,$ and let the coupling energy be of the form $E d$ (where $d$ is the atomic dipole moment); then, in the rotating-wave approximation, the interaction Hamiltonian can be written as $\hbar g(a_\text{cav}\sigma^\dagger + a^\dagger_\text{cav}\sigma)$, where the coupling constant $g= (d^2\Omega_0/\hbar\epsilon_0 A l)^{1/2}$, $\sigma^\dagger$ is an atomic raising operator, and the intracavity effective quasi-mode field operator is 
\begin{equation}
a_\text{cav} = \sum_k \frac {\sqrt{2\kappa c/L}}{\kappa -i(\Omega_k - \Omega_c)}\left(\tau_1 a_k+\tau_2 a_{-k}\right) 
\label{e9}
\end{equation}
One can verify that $[a_\text{cav},a_\text{cav}^\dagger] =1$, using the canonical commutation relations for the $a_k, a_{-k}$ modes, and converting the sum over $k$ to an integral via $\sum_k \to 2\pi/L \int dk$; here $2\pi c/L$ is the frequency spacing appropriate for running waves in a cavity of length $L$ (with periodic boundary conditions), as opposed to $\pi c/L$ for the standing-wave quantization assumed in \cite{prev}.

The crucial point of Eq.~(\ref{e9}) is that, in general, the field that couples to the atom is a superposition of modes incident on the cavity \emph{from both sides}.  Defining ``coupled mode operators'' $a_{ck} =  \tau_1 a_{k}+\tau_2 a_{-k}$ and ``uncoupled mode operators'' $a_{uk} =  \tau_2 a_{k}-\tau_1 a_{-k}$, it is clear that if the cavity is driven only from, say, the left side by a one-photon pulse, the incident state is a superposition of the form $\tau_1 \ket{1}_c + \tau_2 \ket{1}_u$, where the second term does not participate in the interaction at all.  If our goal is, for instance, to change the state of the atom (as with a SWAP gate) this immediately leads to a failure probability of at least $\tau_2^2$.


Essentially, ``opening'' the other side of the cavity turns a scheme for quantum logic that was originally deterministic into a probabilistic one.  The damage can be quantified in the following way: in Eq.~(\ref{n1}), take all the photon states on the left-hand side to be left-incident single-photon states, and write them as a superposition of coupled and uncoupled modes, as above; apply the transformation (\ref{n1}) to the coupled part only; then rewrite the final result in terms of left- and right-incident modes again.  The ``error'' terms are all those containing a factor of $\tau_2$.  

These results highlight an important difference between the KIN scheme and the Duan-Kimble-type gates.  The latter always work in the limit of negligible atomic excitation, in which the atom-cavity system acts as a whole, imparting a phase change to the photon without changing its own state. In the KIN scheme, instead, it is essential that the photon couple strongly enough to the atom to change its state, and in general it is impossible for a single photon coming from a definite direction to deterministically change the state of an atom that is coupled just as strongly to a vacuum mode traveling in the opposite direction.  The same problem would be encountered, therefore, in ring cavities such as microdisks or microtoroids.  

In principle, however, as long as the cavity is lossless a way around the difficulty exists.  One must, in the first place, drive the cavity from \emph{both} ends in such a way as to excite only the coupled mode, and then collect all the light leaving the cavity and recombine it appropriately.

A possible arrangement is shown in Figure 3. For optimal coupling, the beam splitter should transmit mode $a^\prime$ with amplitude $\tau_1$, and reflect it with amplitude $\tau_2$.  Then, as shown in the figure, one has $a_k=\tau_1 a^\prime -\tau_2 a^{\prime\prime}$, and $a_{-k} = \tau_1 a^{\prime\prime} +\tau_2 a^\prime$ (where $a^{\prime\prime}$ is a mode in the vacuum state), and the ``coupling mode'' combination $a_{ck} =  \tau_1 a_{k}+\tau_2 a_{-k} \equiv a^\prime$.  Hence a photon entering the beam splitter at port $a^\prime$ goes whole into the coupling mode.  

   \begin{figure}
   \includegraphics[width=8cm]{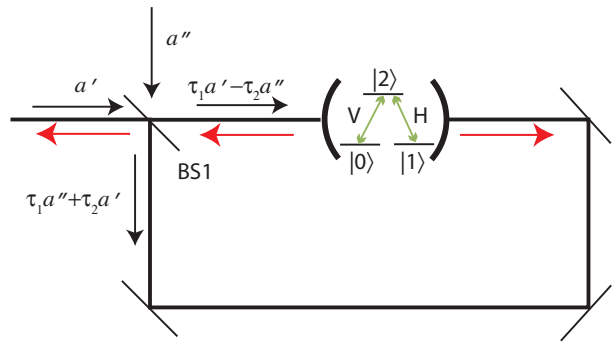}
  \caption[example] 
   { \label{fig:fig3} 
(Color online) Optimal coupling to the double-sided cavity (see text for details).  The lighter arrows represent the outgoing field.  The incoming field $a^{\prime\prime}$ is in the vacuum state.}
   \end{figure} 

The atom also emits the photon in the coupling mode, so its fate upon reaching the beamsplitter can be determined by considering the linear superposition of the four waves $A$, $D$, $A'$ and $D'$ in Figure 2, with the precise phase relationship corresponding to that mode.  Note that in our solution we have assumed, as shown in Fig.~1, that the incident field at the right mirror and the incident field at the left mirror have the same phase, so in Figure 2 with $\tau_1$ and $\tau_2$ real and positive the optical path lengths from the beamsplitter to either mirror must be the same.  Then the left-traveling field arriving at the beam splitter by the upper path is proportional to $\tau_1 A + \tau_2 D^\prime$, and the one arriving from below is proportional to $\tau_1 D+\tau_2 A^\prime$, times a common phase factor.  Near resonance we have
\begin{align}
A = -A^\prime &\simeq \frac{r_2-r_1}{1-r_1r_2} \simeq \frac{t_1^2-t_2^2}{t_1^2+t_2^2} =\tau_1^2 -\tau_2^2 \cr
D = D^\prime &\simeq \frac{t_1 t_2}{1-r_1r_2} \simeq \frac{2 t_1 t_2}{t_1^2+t_2^2}= 2\tau_1\tau_2
\label{e11}
\end{align}
Using these expressions, we find that the left-traveling and upward-traveling fields at the beamsplitter go as $\tau_1$ and $\tau_2$, respectively, times a common phase factor.  Combining them, we see that \emph{no} field leaves the beamsplitter in the upwards direction (since the reflection coefficient on that side of the mirror must be equal to $-\tau_2$).  Hence, in this arrangement, a photon emitted in the coupling mode must leave the beam-splitter along the same path (but in the opposite direction) as the input field.    

We conclude that the arrangement in Fig. 2 allows one to couple optimally to the cavity-atom system, for both input and output.  A similar arrangement would be possible for a ring cavity, for which the coupling mode would be a standing wave; as in Fig.~2, in that case also one should use a beamsplitter to feed the input field into both counterpropagating modes simultaneously, and to collect the output field.

\section{Non-degenerate ground state}

Returning to Eq.~(\ref{e9}), suppose that either the cavity is one-sided, so $\tau_2=0$, or we arrange to excite, and collect, only the coupled mode.  In either case we can then just use the symbol $a_k$ for that mode's component at frequency $\Omega_k = ck$.  Introducing the two polarizations $H$ and $V$, and an atom with two corresponding transitions, we find, in the interaction picture, the Hamiltonian for the system:
\begin{align}
H =  \hbar g \sum_k & \frac{\sqrt{2 c\kappa/L}}{\kappa - i(\Omega_k-\Omega_c)}\, \Bigl(a_{hk} \ket 2 \bra{1} e^{- i(\Omega_k - \omega_{0h}) t} \cr
&+ a_{vk} \ket 2 \bra{0} e^{- i(\Omega_k - \omega_{0v}) t} \Bigr)+ \text{H.c.}
\label{n12}
\end{align}
where the atomic states are labeled as in figures 1 and 3 and $\omega_{0h}$ and $\omega_{0v}$ are the corresponding resonant frequencies.  This Hamiltonian differs from the one we used in \cite{prev} in only a few particulars: the quantization length/frequency spacing (already mentioned above), a phase factor which is discussed in more detail in the Appendix, and the possibility of nondegenerate atomic transitions, with which this section is concerned.

The original KIN proposal assumes that the two ground-state sub levels that make up the atomic qubit are degenerate. This may not always be the case; for instance, for multilevel atoms one might want to deliberately break up the degeneracy of the various magnetic sublevels in order to confine the evolution to an appropriate manifold.  Other systems that have been proposed for quantum logic in a solid-state setting explicitly start out with two non-degenerate lower levels: for instance, doubly excited quantum dots, or nitrogen vacancy centers in diamond.  

Assuming the state vector is of the form
\begin{align}
\ket{\Psi(t)} = &\sum_k C_{hk}(t)\ket 1 a_{hk}^\dagger\ket{\text{vac}} + \sum_k C_{vk}(t)\ket 0 a_{vk}^\dagger\ket{\text{vac}} \cr &
+ C_e(t)\ket 2 \ket{\text{vac}} 
\label{n13}
\end{align}
(where $\ket{\text{vac}}$ is the field vacuum state) we obtain the equations of motion
\begin{subequations}
\begin{align}
\dot C_{hk} &= -ig \frac{\sqrt{2c\kappa/L}}{\kappa + i(\Omega_k-\Omega_c)}\,e^{ i(\Omega_k - \omega_{0h}) t} C_e(t) \label{n14a}\\
\dot C_{vk} &= -ig \frac{\sqrt{2c\kappa/L}}{\kappa + i(\Omega_k-\Omega_c)}\,e^{ i(\Omega_k - \omega_{0v}) t} C_e(t) \label{n14b}\\
\dot C_e &= -ig \sum_k \frac{\sqrt{2c\kappa/L}}{\kappa - i(\Omega_k-\Omega_c)}\, \Bigl( C_{hk} e^{- i(\Omega_k - \omega_{0h}) t} \cr &\qquad\qquad + C_{vk} e^{- i(\Omega_k - \omega_{0v}) t} \Bigr) \label{n14c}
\end{align}
\label{n14}
\end{subequations}
We shall begin by assuming that the pulse is horizontally polarized and the atom in the state $\ket 1$.  We can integrate formally Eqs.~(\ref{n14a}) and (\ref{n14b}), substitute in (\ref{n14c}), and carry out the sum over $k$ in the continuum limit, as we did in \cite{prev}.  The result is the integral equation
\begin{align}
\dot C_e = &-g^2 \int_0^t \left(e^{-(\kappa +i\delta_h)(t-t^\prime)}+ e^{-(\kappa +i\delta_v)(t-t^\prime)}\right) C_e(t^\prime) dt^\prime  \cr
&- ig \int d\omega \frac{\sqrt{\kappa/\pi}}{\kappa - i\omega}\,  C_h(\omega,0) e^{- i(\omega + \delta_a) t}
\label{n15}
\end{align}
where the atom-cavity detunings $\delta_h = \Omega_c-\omega_{0h}$ and $\delta_v = \Omega_c-\omega_{0v}$  have been introduced, as well as the continuous version of the mode coefficients: $C_h(\omega,t) = C_{hk}(t)\sqrt{L/2\pi c}$ (this is appropriate for a mode spacing $2\pi c/L$ and ensures $\int |C_h(\omega)|^2 d\omega = \sum_k |C_{hk}|^2 = 1$).  The frequency $\omega$ is taken to be zero at the cavity resonance.  The advantage of the continuous formulation is that it leads to a straightforward solution of (\ref{n15}) (something which we failed to appreciate in \cite{prev}), in the frequency domain.  Introducing the Fourier transform of $C_e(t)$ by
\begin{equation}
C_e(t) = \frac{1}{2\pi}\int \tilde C_e(\omega) e^{-i\omega t} d\omega
\label{n16}
\end{equation}
it is easy to see that Eq.~(\ref{n15}) implies
\begin{align}
\tilde C_e(\omega) &= -2ig \sqrt{\frac{\pi\kappa}{c}}\times \cr
&\frac{C_h(\omega-\delta_h, 0)}{g^2(2+i\epsilon/(\kappa-i(\omega+\delta_v)))-i\omega(\kappa-i(\omega-\delta_h))} \cr
\label{n17}
\end{align}
where $\epsilon \equiv \delta_h-\delta_v$. Also, the time integrals of the right-hand sides of Eqs.~(\ref{n14a}) and (\ref{n14b}), from well before the interaction starts (which may alternatively be considered the time $0$, or $-\infty$) to well after it is over (formally plus infinity) are then just proportional to $\tilde C_e(\omega+\delta_h)$ and $\tilde C_e(\omega+\delta_v)$, respectively.  Putting this together with Eq.~(\ref{n17}) immediately yields the frequency spectra of the outgoing horizontally and vertically polarized pulses:
\begin{widetext}
\begin{subequations}
\begin{align}
C_h(\omega,\infty) &= C_h(\omega,0) - \frac{2 g^2\kappa}{\kappa + i\omega}\,\frac{C_h(\omega, 0)}{g^2(2+i\epsilon/(\kappa-i(\omega+\epsilon)))-i(\omega+\delta_h)(\kappa-i \omega)}
\label{n18a}
\\
C_v(\omega,\infty) &= - \frac{2 g^2\kappa}{\kappa + i\omega}\,\frac{C_h(\omega-\epsilon, 0)}{g^2(2+i\epsilon/(\kappa-i\omega))-i(\omega+\delta_v)(\kappa-i(\omega-\epsilon))}
\label{n18b}
\end{align}
\label{n18}
\end{subequations}
\end{widetext}
These expressions show that the outgoing vertical pulse is centered at a frequency different from the one of the incoming horizontal pulse, as one would expect from conservation of energy (see Fig. 4).

Leaving the question of this frequency shift aside for a moment, we may consider what is required in order to get the ideal transformation indicated in Eqs.~(\ref{n1}). Ideally, the final pulse should be a perfect replica of the initial one, which would only be possible if the coefficients multiplying $C_h(\omega,0)$ in Eqs.~(\ref{n18}) did not depend on $\omega$.  This is clearly not the case, but for a very long pulse, centered around some frequency $\Omega_h$, the spectrum may be so sharp that it is a good approximation to evaluate those factors only at that frequency.  This is essentially the adiabatic approximation, and departures from it will be considered in the next section.  

For the remainder of this section, therefore, we will treat the prefactors in Eqs.~(\ref{n18}) as constants evaluated at the frequency where $C_h(\omega,0)$ is maximum.  Note that $\omega$ in these expressions is defined relative to the cavity frequency, so if the incoming pulse peaks at $\Omega_h$, the corresponding value of $\omega$ in Eq.~(\ref{n18a}) is $\Omega_h - \Omega_c \equiv\Delta_h$; also, in Eq.~(\ref{n18b}), $C_h(\omega-\epsilon,0)$ peaks at $\omega = \Delta_h + \epsilon$, so it is at this frequency that the coefficient should be evaluated.  Under these conditions, therefore, Eqs.~(\ref{n18}) yield the approximate transformation
\begin{equation}
\ket{H,1} \to \alpha \ket{H,1} + \beta\ket{V,0}
\label{n19}
\end{equation}
with 
\begin{align}
\alpha &=  1 - \frac{2 g^2\kappa}{\kappa + i\Delta_h} \times \cr
&\,\frac{1}{g^2(2+i\epsilon/(\kappa-i(\Delta_h+\epsilon)))-i(\Delta_h+\delta_h)(\kappa-i \Delta_h)} \cr
\beta &= -\frac{\kappa + i\Delta_h}{\kappa + i(\Delta_h+\epsilon)}(1-\alpha)
\label{n20}
\end{align}
It is now an easy matter to see that, in the degenerate case ($\epsilon=0$), the condition (\ref{n2}) does indeed produce a $\sqrt{\text{SWAP}}$ gate, that is, Eq.~(\ref{n1b}), with $\phi=\pi/4$.  On the other hand, in the nondegenerate case one can still make the coefficient $\alpha$ equal to the desired value, $(1-i)/2$, provided the detunings are chosen to satisfy 
\begin{subequations}
\begin{align}
\Delta_h &= -\frac{\epsilon}{2} \label{e23a} \\
\delta_h &= \frac{\epsilon}{2} + \frac{2g^2}{\kappa(1+\epsilon^2/4 \kappa^2)}
\label{e23b}
\end{align}
\end{subequations}
(note this solution is unique). In that case one also has
\begin{equation}
\beta = -\frac{\kappa - i\epsilon/2}{\kappa + i\epsilon/2}\,\frac{1+i}{2} = -e^{-i\theta} \,\frac{1+i}{2}
\label{n22}
\end{equation}
which is also equal to the desired value, except for a phase factor $e^{-i\theta}$, with $\theta = \tan^{-1}(\epsilon/\kappa)$.  As we shall see below, this does not necessarily spoil the $\sqrt{\text{SWAP}}$ gate operation.

We still need to check the third of Eqs.~(\ref{n1}) (the first and fourth are, of course, trivial).  For this the initial state should be chosen to be $\ket{V,0}$, and, for consistency, the vertically-polarized pulse should be chosen to be centered at $\Omega_v = \Omega_h + \epsilon$, as indicated above.  In that case, all the math leading to Eqs.~(\ref{n18}) still holds, only with $v$ and $h$ interchanged, and the sign of $\epsilon$ reversed.  Now the choice $\Delta_h = -\epsilon/2$ implies $\Delta_v = \Omega_h + \epsilon -\Omega_c = \Delta_h + \epsilon = \epsilon/2$, and we obtain
\begin{equation}
\ket{V,0} \to \frac{1-i}{2} \ket{V,0} - e^{i\theta}\,\frac{1+i}{2}\ket{H,1}
\label{n23}
\end{equation}
which is again the desired result except for a phase factor.  The form of this factor shows that, indeed, the operation is unitarily equivalent to a $\sqrt{\text{SWAP}}$, since it is the opposite of the one appearing in front of $\ket{V,0}$ in  the transformation (\ref{n19}) (according to Eq.~(\ref{n22})).  In practice, one could eliminate these factors by multiplying the $H$ component of the incoming pulse by $e^{i\theta}$ before the interaction, and by $e^{-i\theta}$ afterwards.

In short: the $\sqrt{\text{SWAP}}$ gate can be performed (in the adiabatic approximation) exactly provided the vertically and horizontally polarized pulses have, from the start, different frequencies.  This frequency difference is then preserved throughout the interaction. The precise detuning condition needed is illustrated in Fig.~4.

   \begin{figure}
   \includegraphics[width=7cm]{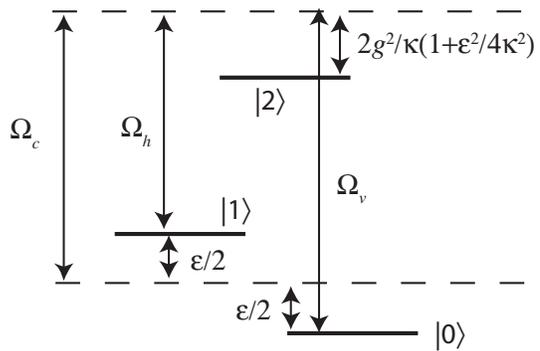}
  \caption[example] 
   { \label{fig:fig4} 
Detunings necessary for a $\sqrt{\text{SWAP}}$ gate.  The transition $\ket 0 \to \ket 2$ couples to vertically-polarized radiation with frequency $\Omega_v$, and the transition $\ket 1 \to \ket 2$ to horizontally-polarized radiation with frequency $\Omega_h$.  The resonant frequency of the cavity is $\Omega_c$.}
   \end{figure} 

While this sounds good, the problem with having different frequencies for the two polarizations  is that single-qubit gates---which in the degenerate case were simple polarization rotations---now become difficult to implement.  A 45 degree rotation applied to both fields, for instance, creates a superposition of fields with different frequencies at each polarization component.  So for a single-qubit gate one would have to first separate the polarizations, frequency-shift them so that they both have the same frequency, combine them and rotate them, separate them again and frequency-shift them back to what they were, and finally combine them again.

Alternatively, as long as one is only doing single-qubit operations, one could just keep the two polarizations degenerate, and only up- or down-shift them before they go into the cavity, and after they come out.  We note that frequency shifting of single photons that preserves polarization entanglement has been recently demonstrated experimentally \cite{zeilinger}, so this is definitely feasible, in principle.

For small $\epsilon$ it may be tempting to ignore the frequency shifts altogether and work with (nominally) degenerate polarizations anyway.  In this case, there is a dual price to pay: a mis-overlap of the pulses in frequency space after each interaction with the cavity, which can be estimated as $\sim (\epsilon T)^2$  for a pulse of duration $T$, and an error in the gate itself.  Both of these can be calculated together from the exact Eqs.~(\ref{n18}), although the gate error may also be estimated from the approximate Eqs.~(\ref{n19}) and (\ref{n20}). For instance, suppose one simply tunes both $\Omega_h$ and $\Omega_v$ to the cavity resonance $\Omega_c$, and sets the latter so that $\delta_h$ and $\delta_v$ have their correct values, as in Fig. 4 and Eq.~(\ref{e23b}).  Then the error in the coefficients $\alpha$ and $\beta$ goes as
\begin{equation}
\left(-\frac{1}{4\kappa} + \frac{\kappa}{8 g^2} \right) \epsilon
\label{e30}
\end{equation}
to first order in $\epsilon$.  Interestingly, this first-order error can be made to vanish by using a cavity that satisfies $g = \kappa/\sqrt 2$.  Nonetheless, even under these conditions, one still finds a second-order gate error of the order of $(\epsilon/\kappa)^2$.  Recalling that the adiabatic condition requires $\kappa T \gg 1$, we see that the gate error will typically be smaller than the  frequency mis-overlap error; that is, the basic criterion for the frequency split $\epsilon$ to be negligible is
\begin{equation}
\epsilon^2 T^2  \ll 1
\label{e31}
\end{equation}
which may be difficult to satisfy in the adiabatic regime. The requirement makes sense, physically, since in this regime the pulse duration provides the smallest bandwidth in the system, hence the pulse itself (not the cavity) is the sharpest frequency discriminator.

\section{Corrections to the adiabatic approximation}

From all of the above we see that it is important to try to determine how long the pulse has to be, in practice, before the results obtained under the adiabatic approximation cease to be valid.  Put otherwise, we want to calculate the leading corrections to the ``very long pulse approximations'' that were made in \cite{prev}.

This is actually not difficult, since we have the exact solution (\ref{n18}) in the frequency domain, valid for any initial pulse.  For simplicity, in the remainder of this section we will be concerned only with the degenerate case, $\epsilon = 0$, and we will simply write $\delta_a$ for the atom-cavity detuning $\Omega_c - \omega_0$.  Then Eqs.~(\ref{n18}) yield
\begin{subequations}
\begin{align}
C_h(\omega,\infty) &= \frac{2 i g^2\omega-i(\omega+\delta_a)(\kappa^2 + \omega^2)}{2 g^2(\kappa+i\omega)-i(\omega+\delta_a)(\kappa^2 + \omega^2)}C_h(\omega,0)
\label{n26a}
\\
C_v(\omega,\infty) &= - \frac{2 g^2\kappa}{2 g^2(\kappa+i\omega)-i(\omega+\delta_a)(\kappa^2 + \omega^2)} C_h(\omega,0)
\label{n26b}
\end{align}
\label{n26}
\end{subequations}
In our previous paper \cite{prev}, we introduced coefficients $C_{k+} = (C_{hk} + C_{vk})/\sqrt 2$, in terms of which a fidelity $F$ was defined by
\begin{align}
e^{2i\Phi}F &= -2 \sum_k C_{k+}(0)^\ast C_{k+}(\infty) \cr
&= -2\int C_+(\omega,0)^\ast C_+(\omega,\infty)\, d\omega
\label{n27}
\end{align}
where, in the second line, we have expressed the same quantity in terms of the continuous-spectrum coefficients we are using here.  It is easy to see, using Eqs.~(\ref{n26}), that this expression is given by
\begin{align}
e^{2i\Phi}F &= \int \frac{2 g^2(\kappa -i\omega)+i(\omega+\delta_a)(\kappa^2 + \omega^2)}{2 g^2(\kappa+i\omega)-i(\omega+\delta_a)(\kappa^2 + \omega^2)}|C_h(\omega,0)|^2 \, d\omega \cr
&= \int e^{2i\psi(\omega)} |C_h(\omega,0)|^2\, d\omega
\label{n28}
\end{align}
with
\begin{equation}
\psi(\omega) = \tan^{-1}\left[\frac{(\omega+\delta_a)(\kappa^2 + \omega^2)-2g^2\omega}{2g^2\kappa}\right]
\label{n29}
\end{equation}
Clearly, in the adiabatic limit, when $C_h(\omega,0)$ is so sharply peaked that $\omega$ can be replaced by a constant detuning, $\Delta$, in the phase factor in the integrand, one has $F=1$ and $\Phi = \psi(\Delta)$.  Additionally, in this case, as it is easy to see, the phase $\Phi$ equals the parameter $\phi$ appearing in the ideal transformation (\ref{n1}) (compare also Eq.~(\ref{n29}) to Eq.~(34) of \cite{prev}).

For shorter pulses, one may write $\omega = \Delta + \omega^\prime$, expand (\ref{n29}) around $\omega = \Delta$, and substitute in (\ref{n28}).  Assuming a Gaussian pulse, $|C_h(\omega,0)|^2 =(T/\sqrt{2\pi}) \exp[-(\omega-\Delta)^2 T^2/2]$ (compare Eq.~(12) of \cite{prev}), after some straightforward algebra, one obtains the following  relatively compact form for the fidelity:
\begin{align}
F^2 = &1 - \frac{16 g^4 \kappa^2 \bigl(-2g^2+2\delta_a\Delta+3\Delta^2 + \kappa^2\bigr)^2}
{T^2\bigl(4g^4\kappa^2+((\Delta+\delta)(\Delta^2+\kappa^2)-2g^2\Delta)^2\bigr)^2} \cr
&+ O\left(\frac{1}{T^4}\right)
\label{e45}
\end{align}
The most interesting feature of this result is that it is possible to make the $1/T^2$ term vanish for a suitable choice of parameters.  For instance, suppose we want to carry out a $\sqrt{\text{SWAP}}$ gate.  As indicated above, this means we want $\psi(\Delta) = \pi/4$, from which Eq.~(\ref{n2}) immediately follows.  Additionally, in order to make the second-order term in (\ref{e45}) vanish, one should set
\begin{equation}
\delta_a = \frac{2 g^2 - 3\Delta^2 -\kappa^2}{2\Delta}
\label{e47}
\end{equation}
Combining (\ref{n2}) and (\ref{e47}) one gets the following equation for $\Delta$:
\begin{equation}
\Delta^4 + 2 \Delta^2(g^2 + \kappa^2) + 4 g^2 \kappa \Delta + \kappa^2 (\kappa^2 - 2 g^2) = 0
\label{e48}
\end{equation}
It is possible to show that at least one real root to (\ref{e48}) always exists, provided 
\begin{equation}
2 g^2 \ge (12\sqrt 3 -20)\kappa^2 \approx 0.785 \kappa^2
\label{e49}
\end{equation}
that is, provided the cavity is ``good enough''.  

The simplest solution to the equations (\ref{e47}) and (\ref{e48}) happens when $2g^2 = \kappa^2$ exactly.  In that case, one can simply set $\Delta = 0$ and $\delta_a = \kappa$.  However, as long as one can adjust the detunings $\Delta$ and $\delta_a$ independently, such precise engineering of the coupling to the cavity is not necessary: one only has to make sure that $g$ is large enough to satisfy Eq.~(\ref{e49}). 

The possibility of eliminating the error to second order in $1/T$ allows for a substantial relaxation of the pulse length requirements.  We illustrate this with the results of ``exact'' numerical calculations such as the ones reported in \cite{prev}.  Figure 5 shows the behavior of $1-F^2$ (sometimes called the ``infidelity'') as a function of $\Delta$ for a pulse of length $T = 10/\kappa$, $g^2 = (169/175) \kappa^2$, and $\delta_a$ chosen to satisfy (\ref{n2}) for every value of $\Delta$.  The figure shows a decrease in the infidelity of more than one order of magnitude at $\Delta = \kappa/5$, which is an exact root of (\ref{e48}) for this value of $g$.

   \begin{figure}
   \includegraphics[width=8cm]{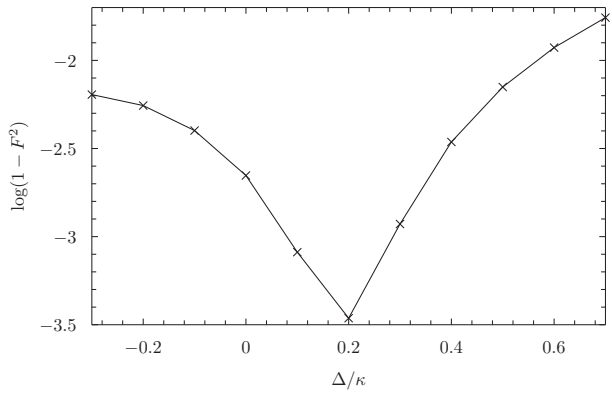}
  \caption[example] 
   { \label{fig:fig5} 
Behavior of the log (base 10) of the infidelity around the point where the second-order term in Eq.~(\ref{e45}) vanishes.  Parameters are: $T=10/\kappa$, $g^2 = (169/175)\kappa^2$, and $\delta_a$ as given by Eq.~(\ref{n2}). The value of $1-F^2$ at the minimum is about $3\times 10^{-4}$.}
   \end{figure} 

Figure 6 shows the dependence of $1-F^2$ on $T$ for $g^2 = \kappa$ and detunings chosen to satisfy Eqs.~(\ref{e47}) and (\ref{e48}).  Both branches of solutions to Eq.~(\ref{e48}) are shown.  The best fit slopes are $-3.95\pm 0.05$ for branch 1 and $-4.03\pm 0.01$ for branch 2.  

   \begin{figure}
   \includegraphics[width=8cm]{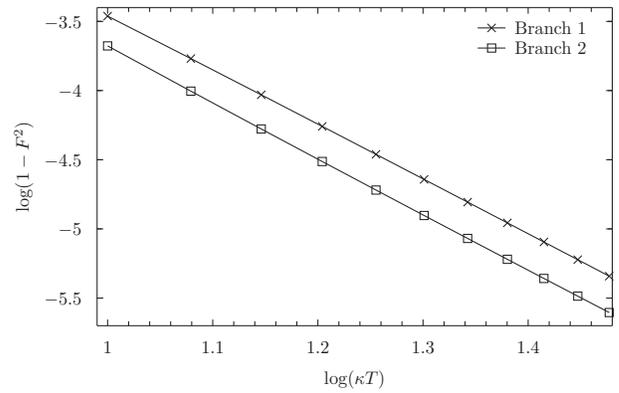}
  \caption[example] 
   { \label{fig:fig6} 
Scaling of the infidelity with the pulse duration for $g = \kappa$.  Both branches of solutions are shown.}
   \end{figure} 

In principle, for an arbitrary initial state, the ``failure probability'' of the gate (defined as 1 minus the absolute value squared of the overlap of the actual final state with the desired final state) will depend both on the deviation of $F$ from unity and on the deviation of the phase $\Phi$ in Eq.~(\ref{n27}) from the desired value $\phi$ (e.g., $\phi=\pi/4$ for a $\sqrt{\text{SWAP}}$ gate).  However, it is easy to see that, since the failure probability has to be a positive quantity by definition, the phase error must enter it squared, to lowest order. Our calculations show that $\Phi-\phi$ scales already as $1/T^2$, so that contribution to the total error is always at least $O(1/T^4)$. 

We can be a bit more precise.  By expanding $\psi(\omega)$ in Eq.~(\ref{n28}) around $\omega=\Delta$, we get, to lowest order in $1/T^2$
\begin{align}
1-F^2 &= \frac{4{\psi^\prime}^2}{T^2}+\ldots \cr
2(\Phi -\phi) &= \frac{\psi^{\prime\prime}}{T^2} +\ldots
\label{e49.2}
\end{align}
where the primes denote derivatives with respect to $\omega$.  By the above argument, it is more important to make $\psi^\prime$ vanish than $\psi^{\prime\prime}$.  We find that, generally, $\psi^{\prime\prime}$ does not vanish for the values of $\delta_a$ and $\Delta$ that cause $\psi^\prime$ to vanish, except, interestingly, at the precise point where the cavity becomes ``good enough'' (according to Eq.~(\ref{e49})), namely, $g \simeq 0.63 \kappa$. This is shown in the numerical calculations plotted in figures 7 and 8.  Except in the neighborhood of this point, however, it appears that  the actual value of $g/\kappa$ does not make much of a difference to either the fidelity or the phase error.  The figures also show that, in general, depending on the values of $T$ and $g/\kappa$, one branch or the other of the solutions of (\ref{e49}) may give a higher fidelity, but the differences are not very dramatic.

   \begin{figure}
   \includegraphics[width=8cm]{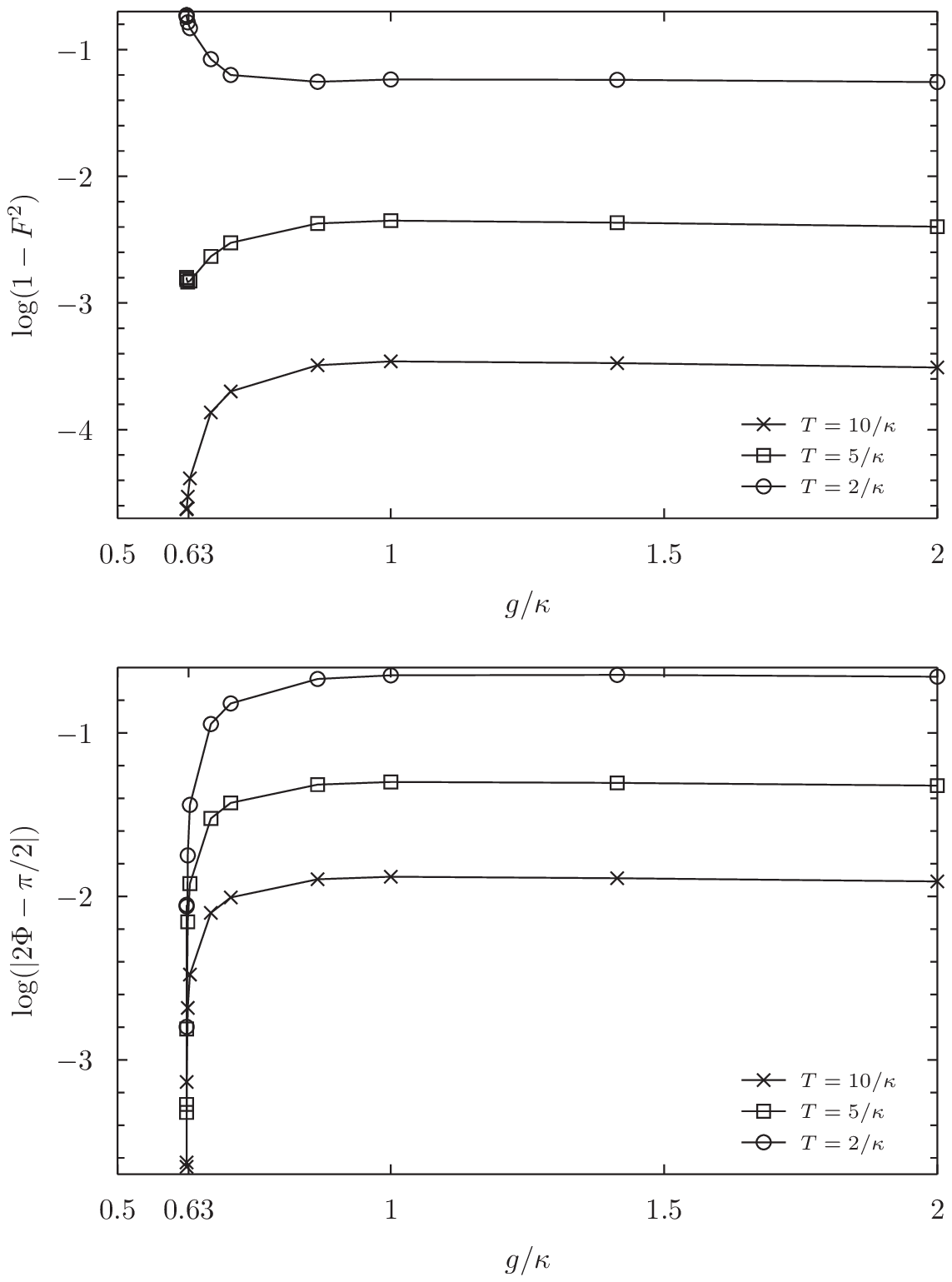}
  \caption[example] 
   { \label{fig:fig7} 
Logarithmic (base 10) plots of the infidelity and the phase error as a function of $g/\kappa$, for the values of the detuning corresponding to one of the branches of solutions of Eq.~(\ref{e48}).  Pulse durations of $T=10/\kappa$, $5/\kappa$ and $2/\kappa$ are shown. }
   \end{figure} 

   \begin{figure}
   \includegraphics[width=8cm]{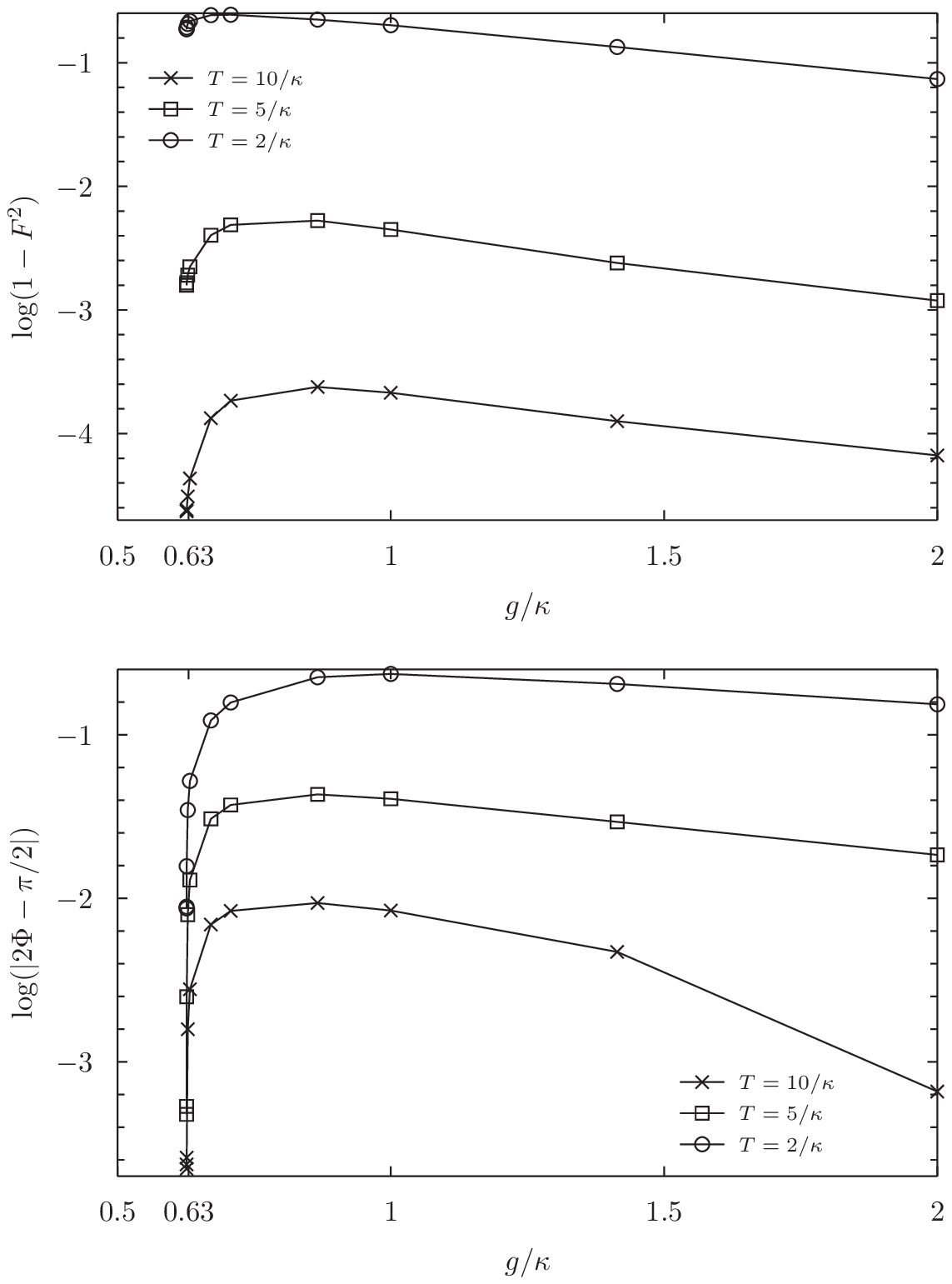}
  \caption[example] 
   { \label{fig:fig8} 
Same as Fig. 7, for the other branch of solutions of Eq.~(\ref{e48}).}
   \end{figure} 

Lastly, for completeness, we would like to briefly discuss the nonadiabatic corrections to
the direct SWAP gate. Although a SWAP between the atom and the photon can always be carried out via two consecutive $\sqrt{\text{SWAP}}$ gates, it may be useful in some cases to perform this operation in a single step, and, as pointed out in the introduction (see also \cite{lin}) the KIN mechanism allows one to do that.  In principle, the simplest way to do it is to set $\Delta = \delta_a =0$, but if one also wants to eliminate the second-order term in (\ref{e45}) an alternative must be sought.  As Eq.~(\ref{n29}) shows, for a SWAP gate one has to impose $(\Delta + \delta_a)(\kappa^2+\Delta^2) - 2 g^2\Delta = 0$, which is satisfied whenever
\begin{equation}
\delta_a = \Delta\left(\frac{2 g^2}{\kappa^2 + \Delta^2} - 1\right)
\label{e50}
\end{equation}
If Eq.~(\ref{e50}) holds, the quadratic term in (\ref{e45}) can be made to vanish by the choice
\begin{equation}
\Delta^2 = g\sqrt{g^2 + 4 \kappa^2} - g^2 - \kappa^2
\label{e51}
\end{equation}
This equation always has a real solution provided that $2g^2 > \kappa^2$ (the good cavity regime again).  Hence, in this regime, high-fidelity single step SWAP gates with relatively short pulses are also possible.  Numerical results for the infidelity are shown in Fig. 9.

   \begin{figure}
   \includegraphics[width=8cm]{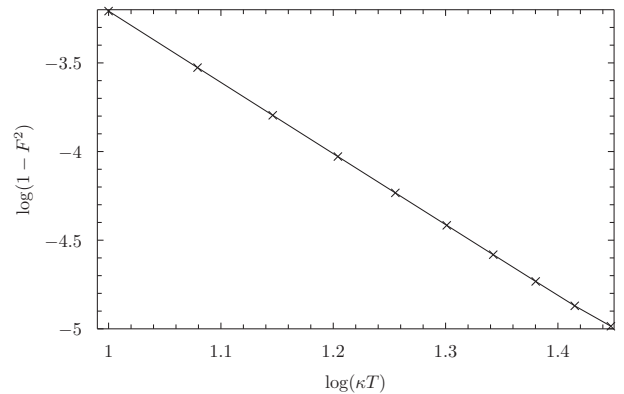}
  \caption[example] 
   { \label{fig:fig9} 
Scaling of the infidelity with the pulse duration for $g = \kappa$ and detunings chosen to satisfy the SWAP condition.}
   \end{figure}

\section{Conclusions}

In this paper, we have studied how to carry out cavity-mediated photon-photon
gates by the KIN mechanism in a variety of systems that depart from the original
proposal in a number of ways: two-sided or bidirectional cavities, nondegenerate atomic
transitions, and somewhat longer pulses. We find that, in principle, all of these systems
are usable, although at the expense of greater experimental complexity. Alternatively, our
results show how the KIN gates deteriorate when the original ideal conditions are not met,
if compensating measures are not taken.

A particularly interesting result is the possibility of choosing the detunings so as to
eliminate the first correction to the adiabatic approximation. Although one still needs to
satisfy the condition $T > 1/\kappa$, this means that fairly high fidelities ($F^2\simeq 0.995$) can be achieved for $T$ as short as $5/\kappa$ and improve rapidly as $T$ increases.
All these results increase our hope that a proof of principle of this scheme for photon-photon
quantum logic may be experimentally achieved in the not too distant future.


\acknowledgments

This research has been supported by the National Science Foundation.

\appendix*

\section{Cavity-induced phase shifts}
 
As mentioned at the beginning of Section IV, the Hamiltonian we use here (Eq.~(\ref{n12})) differs from the one we used in \cite{prev} in a couple of particulars. One is that the quantization length $L$ appears everywhere divided by 2, relative to the way it appeared in \cite{prev}; this simply reflects the fact that for the traveling waves introduced in Section III, the natural boundary conditions are periodic, which lead to a mode spacing $2\pi c/L$, instead of $\pi c/L$ as was the case for the standing waves used in \cite{prev}.  This is merely a formal difference, since $L$ is, after all, arbitrary.

A more subtle difference concerns the coupling coefficients to the individual ``modes of the universe'' inside the cavity, which in this paper's treatment have the form 
\begin{equation}
\frac{\sqrt{2 c\kappa/L}}{\kappa -i(\Omega_k-\Omega_c)} = \frac{\sqrt{2 c\kappa/L}}{\sqrt{\kappa^2 +(\Omega_k-\Omega_c)^2}} e^{i\theta_k}
\label{a1}
\end{equation}
which differs from the form used in \cite{prev} (see, e.g., Eq.~(11) of that paper) by a phase factor $e^{i\theta_k}$, with $\theta_k = \tan^{-1}[(\Omega_k-\Omega_c)/\kappa]$.  This difference actually can be traced back to the phase choice for the incident modes in Section III.  As suggested in Fig.~2, here we are implicitly assuming for the incoming field the form
\begin{equation}
E_\text{in}^{(+)} =  \sum_k \left(\frac{\hbar\Omega_k}{2\epsilon_0 A L}\right)^{1/2}  a_k e^{- i\Omega_k t}
\label{a2}
\end{equation}
whereas in \cite{prev} we had
\begin{equation}
E_\text{in}^{(+)} = \frac{1}{2i} \sum_k \left(\frac{\hbar\Omega_k}{\epsilon_0 A L}\right)^{1/2} \xi_k a_k e^{ikL - i\Omega_k t}
\label{a3}
\end{equation}
Apart from some trivial phases, the key difference is in the factor $\xi_k e^{ikL}$ in Eq.~(\ref{a3}).  By following the arguments given at the end of Section III of \cite{prev}, it can be seen that this combination is essentially equal to $e^{-i\theta_k}$, with $\theta_k$ defined as above.  Hence, in both papers, the key point is that the mode operators inside the cavity are multiplied by a factor of $e^{i\theta_k}$, relative to the incident modes.  We believe the phase choice made in the present paper is the more natural one.  We should note that, in fact, in our previous paper our choice of an initial pulse spectrum (Eq.~(12) of \cite{prev}) did not properly account for the $\xi_k e^{ikL}$ factor.  This is relatively unimportant, since the choice of initial state is also to some extent arbitrary, but it needs to be pointed out.

The phase choice also has consequences for the outgoing modes.  In \cite{prev}, the outgoing field was written as 
\begin{equation}
E_\text{out}^{(+)} = -\frac{1}{2i} \sum_k \left(\frac{\hbar\Omega_k}{\epsilon_0 A L}\right)^{1/2} \xi_k a_k e^{-ikL - i\Omega_k t}
\label{a4}
\end{equation}
and the point was made that the overall phase difference with the incoming field was a factor $-e^{-2ikL} = -e^{2i\theta_k}$ in our current notation.  By expanding the result (\ref{e1}) for $A$ in Section III above, on the other hand, we can see that the outgoing field in this paper can be written as
\begin{equation}
E_\text{in}^{(+)} =  \sum_k \left(\frac{\hbar\Omega_k}{2\epsilon_0 A L}\right)^{1/2} e^{2i\theta_k} a_k e^{- i\Omega_k t}
\label{a5}
\end{equation}
with
\begin{equation}
e^{2i\theta_k} = \frac{\kappa + i(\Omega_k-\Omega_c)}{\kappa - i(\Omega_k-\Omega_c)}
\label{a6}
\end{equation}
This again shows the consistency of the two treatments (except for an overall minus sign which depends, basically, on whether one treats the input mirror as having imaginary transmission coefficients or not), within their different phase conventions: the key thing to note is that going from the input field to the cavity field one gains a factor of $e^{i\theta_k}$, and going from inside the cavity to the output field one gains yet another one.

These cavity-induced modifications of the spectrum are ``built-in'' in the modes of the universe formalism, so that, for instance, for a pulse incident on an empty cavity, the state vector coefficients do not change even though the pulse spectrum does.  It can be argued, therefore, that the measure of fidelity that we have used here, involving only the state vector coefficients, does not really capture all the ways in which the interaction with the atom-cavity system can modify the single-photon pulses.  This is technically true but not likely to be very important in practice, for several reasons.  First, as long as the adiabatic approximation holds, the factor (\ref{a6}) is approximately independent of $k$, and hence in this limit one merely has a constant phase factor that affects all pulses equally upon interacting with a cavity, independently of the pulse polarization or the atomic state (in the nondegenerate case, the phase factors for vertically and horizontally-polarized pulses will be different if they have different central frequencies, but again, as long as they are constant, they can be removed by a method analogous to the one mentioned after Eq.~(\ref{n23}) above). Second, the corrections to the adiabatic regime arising from the factors (\ref{a6}) involve the power series 
\begin{equation}
\tan^{-1}\left[\frac{\Omega_k-\Omega_c}{\kappa}\right] \simeq \frac{\Omega_k-\Omega_c}{\kappa} + O\left(\frac{\Omega_k-\Omega_c}{\kappa}\right)^3
\end{equation}
When this is exponentiated, the first term, being linear in $\Omega_k$, simply amounts to a displacement of the pulse in real space, which, being the same for all initial states (unlike the interaction-induced shifts discussed in Section V) can be easily removed or simply accounted for exactly.  This leaves the next order terms, which are cubic corrections to the phase, and would end up squared as contributions to the error probability, so they go at least like $1/T^6$ (where, as usual, the pulse bandwidth is of the order of $1/T$).

\end{document}